\RequirePackage{fixltx2e}
\documentclass[aps,prb,reprint,floatfix]{revtex4-1}

\usepackage{amsmath,amsfonts,amssymb} %
\usepackage{mathtools} %
\usepackage{wasysym} %
\usepackage{bm} %
\usepackage[bbgreekl]{mathbbol} %
\usepackage{graphicx} %
\usepackage[dvipsnames,table]{xcolor} %
\usepackage{tabularx} %
\usepackage[acronym]{glossaries} %
\usepackage{braket} %
\usepackage{fancyhdr} %
\usepackage{microtype} %
\usepackage{natmove}
\usepackage[byname]{smartref} %
\usepackage[breaklinks, %
colorlinks, linkcolor=Blue, citecolor=Blue, urlcolor=Blue]{hyperref} %
\usepackage[version=3]{mhchem}
\AtBeginDocument{\usepackage{booktabs}} 

\newcommand{\GG}{\mathcal{G}}
\newcommand{\LL}{\mathcal{L}}
\newcommand{\eq}[1]{Eq.~(\ref{#1})} %
\def\be{\begin{equation}} %
\def\ee{\end{equation}} %
\def\bea{\begin{eqnarray}} %
\def\eea{\end{eqnarray}} %

\graphicspath{{pics/}}

\newacronym{QPE}{QPE}{quantum phase estimation} %
\newacronym{VQE}{VQE}{variational quantum eigensolver} %
\newacronym{UCC}{UCC}{unitary coupled cluster} %
\newacronym{QCC}{QCC}{qubit coupled cluster} %
\newacronym{FCI}{FCI}{full configurational interaction} %
\newacronym{CASCI}{CASCI}{complete active space configurational
  interaction} %
\newacronym{JW}{JW}{Jordan--Wigner} %
\newacronym{BK}{BK}{Bravyi--Kitaev} %
\newacronym[longplural={degrees of freedom}, %
firstplural={degrees of freedom (DOF)}, plural={DOF}]{DOF}{DOF}{degree
  of freedom} %
\newacronym[longplural={equations of motion}, %
firstplural={equations of motion (EOM)}, %
plural={EOM}]{EOM}{EOM}{equation of motion} %
\newacronym{PES}{PES}{potential energy surface} %
\newacronym{CI}{CI}{configuration interaction} %
\newacronym{QMF}{QMF}{qubit mean-field} %
\newacronym{SQP}{SQP}{sequential quadratic programming} %
\newacronym{RHF}{RHF}{restricted Hartree--Fock}

\begin{document}

\author{Artur F. Izmaylov} 
\email{artur.izmaylov@utoronto.ca}
\affiliation{Department of Physical and Environmental Sciences,
  University of Toronto Scarborough, Toronto, Ontario, M1C 1A4,
  Canada; and Chemical Physics Theory Group, Department of Chemistry,
  University of Toronto, Toronto, Ontario, M5S 3H6, Canada}

\title{On construction of projection operators}

\date{\today}

\begin{abstract}
The problem of construction of projection operators on eigen-subspaces of symmetry operators is considered.
This problem arises in many approximate methods for solving time-independent and time-dependent 
quantum problems, and its solution ensures proper physical symmetries in development of approximate methods.   
The projector form is sought as a function of symmetry operators and their eigenvalues characterizing the 
eigen-subspace of interest. This form is obtained in two steps: 1) identification of algebraic structures within a 
set of symmetry operators (e.g. groups and Lie algebras), and 2) construction of the projection operators for 
individual symmetry operators. The first step is crucial for efficient projection operator construction 
because it allows for using information on irreducible representations of the present algebraic structure.  
The discussed approaches have promise to stimulate further developments of 
variational approaches for electronic structure of strongly correlated systems and in quantum computing. 
\end{abstract}

\glsresetall

\maketitle

\section{Introduction}

Projection operators on eigen-subspaces of operators commuting with the Hamiltonian
are beneficial for development of various approximate methods of solving both time-independent (TI) and 
time-dependent (TD) Schr\"odnger equations (SEs). The operators commuting with the system Hamiltonian 
are also known as symmetries, and their expectation values are conserved throughout the dynamics.  
In TI-SE, being able to project the Hamiltonian on one of the symmetries irreducible eigen-subspace
ensures that any trial wavefunction that is non-orthogonal to the selected eigen-subspace will have 
a proper symmetry after the variational procedure and projection.
In dynamical problems, approximate Hamiltonians (or Liouvillians) that do not commute with symmetries 
due to introduced approximations can be symmetrically restored by applying the corresponding projectors, 
which enforces the approximate dynamics to respect symmetries of the original exact Hamiltonian.  

The projection becomes crucial if various variable mappings are employed for devising approximations, 
for example fermionic to qubit Jordan-Wigner\cite{Jordan:1928/zphys/631} 
or Bravyi-Kitaev\cite{Seeley:2012/jcp/224109} transformations, in quantum computing,
\cite{McArdle:2018we,Olson:2017ud} or the 
mapping between discrete quantum states and continuous variables in quantum-classical dynamics.\cite{Kim:2008gc}    
Very frequently, introducing approximations in the mapped problem violates proper physical symmetries 
in the original formulation and introducing the projection operators is a straightforward way to restores the 
physical symmetries.\cite{Ryabinkin:2018/cvqe,Ryabinkin:2018gx,Kelly:2012gz}  
 
Recently, projection techniques were also extensively used in developing efficient (e.g. polynomial computational cost) 
methods for treatment of strongly correlated systems.\cite{Scuseria:2011ig,Qiu:2016cf,Jake:2018ka}
In strongly correlated systems, mean-field approaches very frequently undergo spontaneous symmetry 
breaking. Projections techniques can be used either after variational procedures accounting 
for electron correlation (projection-after-variation) or within them (variation-after-projection). In both cases, 
projectors preserve correct physical symmetries in the final wavefunction, moreover, in variation-after-projection, 
they allow the variational procedure to use the variational flexibility more efficiently.\cite{Qiu:2018ei} 

A naive view on the construction of a projector for a symmetry operator $\hat O$ can be using the form  
$\hat P_k = \ket{\phi_k}\bra{\phi_k}$, where $\ket{\phi_k}$ is a corresponding 
eigenfunction $\hat O \ket{\phi_k} = o_k \ket{\phi_k}$. Yet, this approach is not acceptable 
because of at least two difficulties: 1) the $\ket{\phi_k}\bra{\phi_k}$ form may not be compactly 
presentable in the Hilbert space of the problem (e.g. in the coordinate space this operator is 
non-local integral operator), 2) there can be a degenerate subspace corresponding to a single 
eigenvalue $o_k$ containing potentially infinite number of different terms like $\ket{\phi_k}\bra{\phi_k}$.
Thus, it is more prudent to search for the projection operator in the form $\hat P_k = F(\hat O, o_k)$, where 
$F$ is a function of two arguments. The basic principles of construction such functions 
will be discussed in this work.  
 
Another complication in construction of projectors is that, usually, there is not a single operator 
but a whole set of operators $\{\hat O_i\}$ commuting with the Hamiltonian, 
$[\hat H,\hat O_i] = 0$. However, in general, these operators do not commute 
with each other $[\hat O_i,\hat O_j]\ne 0$, and thus do not have a common set of eigenfunctions.
Therefore, the projection using all known symmetries cannot be simply a projection 
on an eigen-subspace of a particular operator.   
The most natural algebraic structure for this set of operators is the Lie algebra.\cite{Fuchs,Gilmore:2008} 
This is quite intuitive 
if we consider that any commutator of two operators, $[\hat O_i,\hat O_j]$, is again an operator 
commuting with the Hamiltonian. Thus, if an initial set of operators commuting with $\hat H$
is known, this set can be extended by forming all possible commutators 
\bea\label{eq:LA}
[\hat O_i,\hat O_j] = \sum_k c_{ij}^{(k)} \hat O_k, 
\eea
here $c_{ij}^{(k)}$ are some constants. An extended set of operators $\{\hat O_k\}$ 
that is closed with respect to the commutation operation in \eq{eq:LA} forms a Lie algebra. 
This Lie algebra will consists of all operators commuting with the Hamiltonian.\footnote{Note that 
formally powers of $\hat O_i^n$ are not defined in the Lie algebra, therefore,
it is not possible to build Taylor series expandable functions for the operators $f(\hat O_i)$, 
such functions would also commute with $\hat H$ 
and could be added to the Lie algebra based on the commutation property. 
Products $\hat O_i^n$ are only defined in the universal enveloping Lie algebra.}

To impose symmetries, the Lie algebraic structure suggests to construct projectors 
on irreducible representations of the Lie algebra. This process is well described in 
mathematical literature on classification of Lie algebras.\cite{Barut,Fuchs,Gilmore:2008} 
Here, I will provide only a basic 
summary and examples while referring the interested reader to more specialized literature.     
One of the essential differences from other works in the literature on symmetry operators 
will be considering not only irreducible subspaces of Lie algebras
but rather construction of projection operators on these irreducible subspaces 
as functions of the symmetry operators. 

In what follows I present a theory that allows one to account for algebraic properties of a set 
of operators commuting with the Hamiltonian and thus to use all available symmetry 
information to construct projectors on irreducible subspaces of symmetry operators. 
Even though I will consider a set of operators commuting with the Hamiltonian, the methods 
discussed here can be applied for a set of operators not necessarily related to symmetries of 
the Hamiltonian, for example this can be operators commuting with the Liouvillian.        

\section{Theory}
\label{sec:theory}

First, treatment of algebraic properties of the symmetry operator set will be considered. 
Chemists, of course, are more familiar with the situation when $\{\hat O_i\}$ form a finite group, 
$\hat O_i\hat O_j = \hat O_k$, as in the case of point group symmetries. However, I would like 
to argue that this group structure is only a useful addition to the more general underlining 
Lie algebraic structure. Nevertheless, it is convenient to separate two cases based on whether 
$\{\hat O_i\}$  form a multiplicative group or not. 
It will be shown that in the latter (more general case) 
accounting for the Lie algebraic structure leads to the problem of 
construction projection operators for some subset of symmetry operators (e.g. $\hat S_z$, and $\hat S^2$ 
for the algebra of spin). Thus, possible 
ways for such constructions will be considered next.  
 
\subsection{Projectors for groups}

Here, I consider the case when operators $\{\hat O_k\}$ belong to multiplicative group $G$.
Existence of the group structure allows one to generate projectors on the group irreducible 
representations following the standard procedure\cite{Wigner}
\bea\label{eq:FG}
\hat P_\Gamma = \frac{d_\Gamma}{|G|} \sum_{k=1}^{|G|} \chi_\Gamma^{*}(\hat O_k) \hat O_k, 
\eea 
where $\Gamma$ is the irreducible representation of interest, $d_\Gamma$ is the dimension of $\Gamma$, 
$|G|$ is the number of the group elements, and $\chi_{\Gamma}(\hat O_k)$ are characters for 
the group elements. In this case, one does not need to deal with Lie algebras and projection operators 
are simply expressed as a linear function of all operators forming the group.



\subsection{Projectors for Lie algebras}

In the case when $\{O_i\}$'s only form the Lie algebra, one can obtain a continuous Lie group using 
the exponentiation of the algebra elements (see appendix A for general exposition). 
Then the same standard machinery as in the group case can be used for projection construction. 
This approach can be illustrated on a simple example of a single symmetry Hermitian operator $\hat O$.
Exponentiation of  $\hat O$, $g(\hat O, \phi) = \exp[i\phi\hat O]$, where $\phi\in[0,2\pi)$, allows one 
to create a continuous compact cyclic group $G$ with elements $g(\hat O, \phi)$.  
Then the continuous analogue of \eq{eq:FG} can be written as 
\bea\label{eq:IG}
\hat P_j = \frac{1}{2\pi}\int_0^{2\pi}   e^{i\phi (\hat O-o_j)} d\phi,
\eea
where $o_j$ is a particular eigenvalue of $\hat O$. 
All cyclic groups are abelian and have one-dimensional irreducible 
representations; each irreducible representation is characterized by the eigenvalue $o_j$, 
hence, the characters of the irreducible representations are $\exp[i\phi o_j]$.
 
However, switching from the algebra to the group is not necessary to obtain the projectors 
on the irreducible representations of the algebra. Moreover, historically, Lie algebras were introduced 
to simplify analysis of the irreducible representations of Lie groups.  
In any case, understanding irreducible representations of the symmetry Lie algebra is a necessary 
step for the group pathway (see appendix A) 
and for a simpler method avoiding the group construction, which is described below.

For all simple or semi-simple Lie algebras (e.g. $su(2)$, the electron spin)
\footnote{Construction of irreducible representations for 
other types of Lie algebras, such as abelian and reductive (direct sums of abelian and simple), is somewhat 
simpler because they can be built combining representations found for simple algebras and/or one-dimensional 
representations for abelian algebras.} 
the standard procedure to construct irreducible representations 
is to select maximal commuting sub-algebra ({\it i.e.} the Cartan sub-algebra), 
this sub-algebra will form the maximal set of all mutually commuting operators 
with the Hamiltonian and the corresponding eigen-values represent good quantum numbers, 
while the eigen-functions form the basis for the irreducible representations. 
For the well-known $su(2)$-case, the usual choice of the Cartan sub-algebra is the $\hat S_z$ operator. 
To further characterize the irreducible representations one can use Casimir operators, 
which commute with all elements of the algebra. By Schur's lemma this commutativity 
makes any Casimir operator to be equivalent to the identity multiplied by a constant 
for any irreducible representation. These constants
are eigen-values of Casimir operators on irreducible representations and along with the full set of quantum numbers 
fully characterize the basis of irreducible representations. In the $su(2)$-case, $\hat S^2$ is the Casimir 
operator an its eigenvalue $S(S+1)$ along with that for $\hat S_z$, $M=-S,...,S$, fully characterize the basis 
of all irreducible representations. Thus, to construct projectors on the basis states of 
irreducible representations it is enough to construct projectors on eigenstates of 
all operators of the Cartan sub-algebra and Casimir operators, I will refer to these 
operators as the fully commuting set.        

For each operator $\hat O_i$ in the fully commuting set, individual projectors for eigen-subspaces 
$\hat P_j^{(i)}$ will be build as a function that depends on $\hat O_i$ and its eigenvalue $o_j^{(i)}$ 
determining the eigen-subspace, $\hat P_j^{(i)} = F(\hat O_i,o_j^{(i)})$.
A total projector on a particular irreducible representation of the Lie algebra 
can be written as $\hat P = \prod_i  F_i(\hat O_i,o_j^{(i)})$,
where the eigenvalues $o_j^{(i)}$ should be chosen so that the projectors in the product are not 
orthogonal to each over (e.g. in the $su(2)$-case, $M\in\{-S,...,S\}$).
The order of the $F_i$ functions does not matter because if $\hat O_i$ operators commute 
their eigen-subspace projectors also commute (see appendix B). 

\subsection{Role of non-commuting elements}

Both algebras and groups, generally contain elements that do not commute. 
It is important to point out a significance of this non-commutativity. It always indicates degeneracy of some 
eigenstates in the Hamiltonian. In other words, if we characterize some basis functions for irreducible 
representations using only commuting sub-algebras or sub-groups, non-commuting parts contain 
information that some of these basis functions form multi-dimensional irreducible representations corresponding 
to the degenerate eigenstates of the Hamiltonian. If it is not for the non-commuting elements, any Hamiltonian
degeneracy would be perceived as completely accidental. Therefore, non-commuting elements providing 
more symmetry related information on the Hamiltonian spectrum, they are accounted in \eq{eq:FG} for groups 
and translated through the Casimir operators in algebras. 

\subsection{Projector as an indicator function of the operator} 

For a single projector $\hat P_j^{(i)}$, $F(\hat O_i,o_j^{(i)})$ can be constructed as 
some differentiable representation of the Kronecker-delta function. The Kronecker-delta
function naturally appears in the spectral decomposition of the projector 
(here, we consider the non-degenerate case, while the degenerate case can be treated 
similarly with more cumbersome notation) 
\bea\label{eq:DK}
\hat P_j^{(i)} &=& \sum_n \ket{\phi_n^{(i)}}\bra{\phi_n^{(i)}} \delta_{nj}\\
&=& \sum_n \ket{\phi_n^{(i)}}\bra{\phi_n^{(i)}} F(x,o_j^{(i)})\vert_{x=o_n^{(i)}},
\eea  
where we substituted the Kronecker-delta function (also known as an indicator function) 
with the differentiable function 
\bea\label{eq:gen}
F(x,o_j^{(i)}) =
\begin{cases}
1,~ x=o_j^{(i)}, \\
0,~ x=o_n^{(i)}, n\ne j, \\
\xi(x) \in [0,1],~ x \ne o_n^{(i)}, \forall n,
\end{cases}
\eea
where $\xi(x)$ can be any smooth function for intermediate values of $x$. 
Due to its differentiability we can expand $F$ in the Taylor series, and this expansion can define 
$F(\hat O_i, o_j^{(i)})$. Then using the Taylor expansion of $F$ and the projector property of 
 $\ket{\phi_n^{(i)}}\bra{\phi_n^{(i)}}^2 = \ket{\phi_n^{(i)}}\bra{\phi_n^{(i)}}$ one can obtain 
\bea
\hat P_j^{(i)} &=& F(\hat O_i, o_j^{(i)}).
\eea
There are multiple ways to define differentiable representation of the Kronecker-delta 
function $F(x,o_j^{(i)})$, here we list several forms:\\
{\it 1) Integration over a unit circle in real space:} 
\bea\label{eq:exp1}
F(x,o_j^{(i)}) = \frac{1}{2\pi}\int_0^{2\pi} e^{i\phi(x-o_j^{(i)})}d\phi
\eea
Here, for any $x\ne o_j^{(i)}$ we obtain zero. Such selectivity comes with a price of introducing
the integral. \\
{\it 2) Integration over a unit circle in the complex plane:}
\bea
F(x,o_j^{(i)}) = \frac{1}{2\pi i} \oint_{|z|=1} z^{(x-o_j^{(i)}-1)} dz
\eea
This function exploits the same idea as the previous one but using the complex plane.\\ 
{\it 3) The Lagrange interpolation product:}  
\bea\label{eq:LIP}
F(x,o_j^{(i)}) = \prod_{n\ne j} \frac {x - o_n^{(i)}}{o_j^{(i)}- o_n^{(i)}}, 
\eea
which is less restrictive since for $x$-values in between the eigenvalues the functional value is not fixed 
to zero or one. This polynomial function is used in the Lagrange interpolation method.\cite{num_rec} 
The substitution, $x\rightarrow \hat O_i$ in \eq{eq:LIP} leads to the L\"owdin projector operator used for 
the spin projection.\cite{Lowdin, Lowdin3}\\   
{\it 4) Integration over an arbitrary contour $C(o_j^{(i)})$ encircling only $z=o_j^{(i)}$ in the complex plane:} 
\bea
F(x,o_j^{(i)}) = \frac{1}{2\pi i} \oint_{C(o_j^{(i)})}  \frac{dz}{x-z}
\eea
This function with $x\rightarrow \hat H$ 
is the resolvent used in investigation of the perturbation series.\cite{Kato,Messiah}
{\it 5) The difference between the limits of logistic functions:}  
\bea\notag
F(x,o_j^{(i)}) &=& \lim_{k\rightarrow\infty} (1+ e^{-k(x-o_j^{(i)}+\epsilon)})^{-1} \\
&-&  \lim_{k\rightarrow\infty} (1+ e^{-k(x-o_j^{(i)}-\epsilon)})^{-1},
\eea
where $\epsilon =\min |o_j^{(i)}-o_{j\pm 1}^{(i)}|/2$. 
These limits of logistic functions correspond to the Heaviside functions.\\
{\it 6) ``Bump" function (or mollifier):}
\bea\notag
F(x,o_j^{(i)}) = 
\begin{cases}
\exp\left[\frac{1}{(x-o_j^{(i)})^2-\epsilon}\right],~ x\in(o_j^{(i)}-\epsilon^{\frac{1}{2}},o_j^{(i)}+\epsilon^{\frac{1}{2}}),\\
0,~x\not\in(o_j^{(i)}-\epsilon^{\frac{1}{2}},o_j^{(i)}+\epsilon^{\frac{1}{2}}).
\end{cases}
\eea
Unfortunately, this last example cannot be extended to $x\rightarrow \hat O_i$ because of 
the branch choice based on $x$-value in its definition.

Equation \eqref{eq:LIP} is especially useful to build projectors for operators with a finite number of 
eigenvalues because then the product contains a finite number of terms. 
Interestingly, for such operators, projectors built based on Eqs.~\eqref{eq:exp1} and \eqref{eq:LIP} 
are the same. This is a consequence of only a finite number of linear independent powers  
for an operator with a finite spectrum. Using the Cayley-Hamilton theorem\cite{Gelfand} one can show that any 
function of such an operator is equivalent to $N-1$ polynomial, where $N$ is the number of eigenvalues. 
 
Another interesting connection can be found between projectors based on \eq{eq:exp1} and generalization 
of the group projector in \eq{eq:FG} to an infinite continuous one-parametric cyclic group in \eq{eq:IG}.  

In \eq{eq:DK}, the spectrum of the symmetry operator is assumed to be discrete, if it is not the case, the 
Kronecker-delta function needs to be substituted by the Dirac-delta function and its 
numerous representations as limits of continuous functions. 

\subsection{Construction of the indicator function using orthogonality}

Another way to present the Kronecker-delta function in \eq{eq:DK} is to build the
$F$ function as an expansion $F(x,y) = \sum_n f_n(x)c_n(y)$ that satisfies the following relations
 \bea
 \hat P_j^{(i)} \ket{\phi_k^{(i)}} &=& F(\hat O_i,o_j^{(i)}) \ket{\phi_k^{(i)}} \\
 &=& \sum_n c_n(o_j^{(i)})f_n(\hat O_i) \ket{\phi_k^{(i)}} \\
 &=& \sum_n c_n(o_j^{(i)})f_n(o_k^{(i)}) \ket{\phi_k^{(i)}} \\
 \sum_n c_n(o_j^{(i)})f_n(o_k^{(i)})  &=& \bra{c(o_j^{(i)})}f(o_k^{(i)})\rangle = \delta_{kj}.
 \eea
 Here vectors $\ket{f(o_k^{(i)})}$ and $\ket{c(o_j^{(i)})}$ are defined as 
 \bea\notag
 \ket{f(o_k^{(i)})} &=& \{f_1(o_k^{(i)}), f_2(o_k^{(i)}), ... f_M(o_k^{(i)})\}, \\\notag
 \ket{c(o_j^{(i)})} &=& \{c_1(o_j^{(i)}), c_2(o_j^{(i)}), ... c_M(o_j^{(i)})\}
 \eea 
 and are orthonormal for all eigenvalues. The natural question is how to choose $f_n(x)$ and $c_n(y)$? 
 One of the choices closely related to the group theory construction of the projectors is to take 
 $f_k(x) = c_k(x) = \exp(i k x)$ and to switch to a continuous version of $k$ with substituting 
 summation by integration 
 \bea
 F(\hat O_i,o_j^{(i)})  = \frac{1}{2\pi}\int_{0}^{2\pi} c_{k}^*(o_j^{(i)})f_{k}(\hat O_i) dk,
 \eea
 we arrived to the projector already introduced in \eq{eq:IG}. Applying further restrictions on operator $\hat O_i$,
 one can generate finite sum expansions for the projector on its eigenspaces. Such restrictions are: {\it equidistant} 
 separation between neighboring eigenvalues and {\it finite} number of eigenvalues. The later condition is less crucial 
 because its violation only leads to projectors that can separate eigen-states within a finite subset. 
 The basic idea of this finite construction is on the following representation of the Kronecker-delta 
 \bea
 \delta_{nm} = \frac{1}{M}\sum_{k=1}^{M} e^{2\pi i k(n-m)/M}.
 \eea  
 If $\hat O_i$ has a finite and equidistant spectrum with the distance between its eigenstates $d$ 
 then the projector can be written as 
  \bea\label{eq:FED}
 F(\hat O_i,o_j^{(i)})  &=& \frac{1}{M}\sum_{k=1}^{M} e^{2\pi i k (\hat O_i-o_j^{(i)})/(dM)} \\
 &=& \frac{1}{M}\sum_{k=1}^{M} c_{k}^*(o_j^{(i)})f_{k}(\hat O_i),
 \eea
 where $c_k(x)=f_k(x) = \exp(2\pi i k x/(dM))$. This approach can be used for the electron spin 
 projection $\hat S_z$ and the number of electrons $\hat N$ operators.  

\section{Conclusions}
\label{sec:conclusions}

I reviewed various approaches to construct projectors on irreducible eigen-subspaces of symmetry operators. 
There are two aspects of this problem: 1) accounting for available algebraic structures of the set of symmetry 
operators, and 2) construction of individual projection operators as functions of symmetry operators and 
their eigenvalues. Two algebraic structures, groups and Lie algebras, were discussed. For both structures 
standard methods of construction of irreducible representations were developed in mathematical literature. 
Knowledge of irreducible representations helps to construct functions of symmetry operators into the 
projection operator for a particular irreducible representation. For Lie algebras, various approaches to construct 
individual projection operators for each symmetry operator were considered. The origin of various projection 
functions was found in variety of representations for the Kronecker-delta function. 

\section*{Acknowledgement}
A.F.I. is grateful to A. V. Zaitsevskii and V. Y. Chernyak for useful discussions.  
A.F.I. acknowledges financial support from the Natural Sciences and
Engineering Research Council of Canada. 


\begin{thebibliography}{24}%
\makeatletter
\providecommand \@ifxundefined [1]{%
 \@ifx{#1\undefined}
}%
\providecommand \@ifnum [1]{%
 \ifnum #1\expandafter \@firstoftwo
 \else \expandafter \@secondoftwo
 \fi
}%
\providecommand \@ifx [1]{%
 \ifx #1\expandafter \@firstoftwo
 \else \expandafter \@secondoftwo
 \fi
}%
\providecommand \natexlab [1]{#1}%
\providecommand \enquote  [1]{``#1''}%
\providecommand \bibnamefont  [1]{#1}%
\providecommand \bibfnamefont [1]{#1}%
\providecommand \citenamefont [1]{#1}%
\providecommand \href@noop [0]{\@secondoftwo}%
\providecommand \href [0]{\begingroup \@sanitize@url \@href}%
\providecommand \@href[1]{\@@startlink{#1}\@@href}%
\providecommand \@@href[1]{\endgroup#1\@@endlink}%
\providecommand \@sanitize@url [0]{\catcode `\\12\catcode `\$12\catcode
  `\&12\catcode `\#12\catcode `\^12\catcode `\_12\catcode `\%12\relax}%
\providecommand \@@startlink[1]{}%
\providecommand \@@endlink[0]{}%
\providecommand \url  [0]{\begingroup\@sanitize@url \@url }%
\providecommand \@url [1]{\endgroup\@href {#1}{\urlprefix }}%
\providecommand \urlprefix  [0]{URL }%
\providecommand \Eprint [0]{\href }%
\providecommand \doibase [0]{http://dx.doi.org/}%
\providecommand \selectlanguage [0]{\@gobble}%
\providecommand \bibinfo  [0]{\@secondoftwo}%
\providecommand \bibfield  [0]{\@secondoftwo}%
\providecommand \translation [1]{[#1]}%
\providecommand \BibitemOpen [0]{}%
\providecommand \bibitemStop [0]{}%
\providecommand \bibitemNoStop [0]{.\EOS\space}%
\providecommand \EOS [0]{\spacefactor3000\relax}%
\providecommand \BibitemShut  [1]{\csname bibitem#1\endcsname}%
\let\auto@bib@innerbib\@empty
\bibitem [{\citenamefont {Jordan}\ and\ \citenamefont
  {Wigner}(1928)}]{Jordan:1928/zphys/631}%
  \BibitemOpen
  \bibfield  {author} {\bibinfo {author} {\bibfnamefont {P.}~\bibnamefont
  {Jordan}}\ and\ \bibinfo {author} {\bibfnamefont {E.}~\bibnamefont
  {Wigner}},\ }\href {\doibase 10.1007/BF01331938} {\bibfield  {journal}
  {\bibinfo  {journal} {Z. Phys.}\ }\textbf {\bibinfo {volume} {47}},\ \bibinfo
  {pages} {631} (\bibinfo {year} {1928})}\BibitemShut {NoStop}%
\bibitem [{\citenamefont {Seeley}\ \emph {et~al.}(2012)\citenamefont {Seeley},
  \citenamefont {Richard},\ and\ \citenamefont
  {Love}}]{Seeley:2012/jcp/224109}%
  \BibitemOpen
  \bibfield  {author} {\bibinfo {author} {\bibfnamefont {J.~T.}\ \bibnamefont
  {Seeley}}, \bibinfo {author} {\bibfnamefont {M.~J.}\ \bibnamefont {Richard}},
  \ and\ \bibinfo {author} {\bibfnamefont {P.~J.}\ \bibnamefont {Love}},\
  }\href {\doibase 10.1063/1.4768229} {\bibfield  {journal} {\bibinfo
  {journal} {J. Chem. Phys.}\ }\textbf {\bibinfo {volume} {137}},\ \bibinfo
  {pages} {224109} (\bibinfo {year} {2012})}\BibitemShut {NoStop}%
\bibitem [{\citenamefont {McArdle}\ \emph {et~al.}(2018)\citenamefont
  {McArdle}, \citenamefont {Endo}, \citenamefont {Aspuru-Guzik}, \citenamefont
  {Benjamin},\ and\ \citenamefont {Yuan}}]{McArdle:2018we}%
  \BibitemOpen
  \bibfield  {author} {\bibinfo {author} {\bibfnamefont {S.}~\bibnamefont
  {McArdle}}, \bibinfo {author} {\bibfnamefont {S.}~\bibnamefont {Endo}},
  \bibinfo {author} {\bibfnamefont {A.}~\bibnamefont {Aspuru-Guzik}}, \bibinfo
  {author} {\bibfnamefont {S.}~\bibnamefont {Benjamin}}, \ and\ \bibinfo
  {author} {\bibfnamefont {X.}~\bibnamefont {Yuan}},\ }\href@noop {} {\bibfield
   {journal} {\bibinfo  {journal} {arXiv.org}\ } (\bibinfo {year} {2018})},\
  \Eprint {http://arxiv.org/abs/1808.10402v1} {1808.10402v1} \BibitemShut
  {NoStop}%
\bibitem [{\citenamefont {Olson}\ \emph {et~al.}(2017)\citenamefont {Olson},
  \citenamefont {Cao}, \citenamefont {Romero}, \citenamefont {Johnson},
  \citenamefont {Dallaire-Demers}, \citenamefont {Sawaya}, \citenamefont
  {Narang}, \citenamefont {Kivlichan}, \citenamefont {Wasielewski},\ and\
  \citenamefont {Aspuru-Guzik}}]{Olson:2017ud}%
  \BibitemOpen
  \bibfield  {author} {\bibinfo {author} {\bibfnamefont {J.}~\bibnamefont
  {Olson}}, \bibinfo {author} {\bibfnamefont {Y.}~\bibnamefont {Cao}}, \bibinfo
  {author} {\bibfnamefont {J.}~\bibnamefont {Romero}}, \bibinfo {author}
  {\bibfnamefont {P.}~\bibnamefont {Johnson}}, \bibinfo {author} {\bibfnamefont
  {P.-L.}\ \bibnamefont {Dallaire-Demers}}, \bibinfo {author} {\bibfnamefont
  {N.}~\bibnamefont {Sawaya}}, \bibinfo {author} {\bibfnamefont
  {P.}~\bibnamefont {Narang}}, \bibinfo {author} {\bibfnamefont
  {I.}~\bibnamefont {Kivlichan}}, \bibinfo {author} {\bibfnamefont
  {M.}~\bibnamefont {Wasielewski}}, \ and\ \bibinfo {author} {\bibfnamefont
  {A.}~\bibnamefont {Aspuru-Guzik}},\ }\href@noop {} {\bibfield  {journal}
  {\bibinfo  {journal} {arXiv.org}\ } (\bibinfo {year} {2017})},\ \Eprint
  {http://arxiv.org/abs/1706.05413v2} {1706.05413v2} \BibitemShut {NoStop}%
\bibitem [{\citenamefont {Kim}\ \emph {et~al.}(2008)\citenamefont {Kim},
  \citenamefont {Nassimi},\ and\ \citenamefont {Kapral}}]{Kim:2008gc}%
  \BibitemOpen
  \bibfield  {author} {\bibinfo {author} {\bibfnamefont {H.}~\bibnamefont
  {Kim}}, \bibinfo {author} {\bibfnamefont {A.}~\bibnamefont {Nassimi}}, \ and\
  \bibinfo {author} {\bibfnamefont {R.}~\bibnamefont {Kapral}},\ }\href@noop {}
  {\bibfield  {journal} {\bibinfo  {journal} {J. Chem. Phys.}\ }\textbf
  {\bibinfo {volume} {129}},\ \bibinfo {pages} {084102} (\bibinfo {year}
  {2008})}\BibitemShut {NoStop}%
\bibitem [{\citenamefont {Ryabinkin}\ \emph {et~al.}(2019)\citenamefont
  {Ryabinkin}, \citenamefont {Genin},\ and\ \citenamefont
  {Izmaylov}}]{Ryabinkin:2018/cvqe}%
  \BibitemOpen
  \bibfield  {author} {\bibinfo {author} {\bibfnamefont {I.~G.}\ \bibnamefont
  {Ryabinkin}}, \bibinfo {author} {\bibfnamefont {S.~N.}\ \bibnamefont
  {Genin}}, \ and\ \bibinfo {author} {\bibfnamefont {A.~F.}\ \bibnamefont
  {Izmaylov}},\ }\href@noop {} {\bibfield  {journal} {\bibinfo  {journal} {J.
  Chem. Theory Comput.}\ }\textbf {\bibinfo {volume} {15}},\ \bibinfo {pages}
  {249} (\bibinfo {year} {2019})}\BibitemShut {NoStop}%
\bibitem [{\citenamefont {Ryabinkin}\ \emph {et~al.}(2018)\citenamefont
  {Ryabinkin}, \citenamefont {Genin},\ and\ \citenamefont
  {Izmaylov}}]{Ryabinkin:2018gx}%
  \BibitemOpen
  \bibfield  {author} {\bibinfo {author} {\bibfnamefont {I.~G.}\ \bibnamefont
  {Ryabinkin}}, \bibinfo {author} {\bibfnamefont {S.~N.}\ \bibnamefont
  {Genin}}, \ and\ \bibinfo {author} {\bibfnamefont {A.~F.}\ \bibnamefont
  {Izmaylov}},\ }\href@noop {} {\bibfield  {journal} {\bibinfo  {journal} {J.
  Chem. Phys.}\ }\textbf {\bibinfo {volume} {149}},\ \bibinfo {pages} {214105}
  (\bibinfo {year} {2018})}\BibitemShut {NoStop}%
\bibitem [{\citenamefont {Kelly}\ \emph {et~al.}(2012)\citenamefont {Kelly},
  \citenamefont {van Zon}, \citenamefont {Schofield},\ and\ \citenamefont
  {Kapral}}]{Kelly:2012gz}%
  \BibitemOpen
  \bibfield  {author} {\bibinfo {author} {\bibfnamefont {A.}~\bibnamefont
  {Kelly}}, \bibinfo {author} {\bibfnamefont {R.}~\bibnamefont {van Zon}},
  \bibinfo {author} {\bibfnamefont {J.}~\bibnamefont {Schofield}}, \ and\
  \bibinfo {author} {\bibfnamefont {R.}~\bibnamefont {Kapral}},\ }\href@noop {}
  {\bibfield  {journal} {\bibinfo  {journal} {J. Chem. Phys.}\ }\textbf
  {\bibinfo {volume} {136}},\ \bibinfo {pages} {084101} (\bibinfo {year}
  {2012})}\BibitemShut {NoStop}%
\bibitem [{\citenamefont {Scuseria}\ \emph {et~al.}(2011)\citenamefont
  {Scuseria}, \citenamefont {Jim{\'e}nez-Hoyos}, \citenamefont {Henderson},
  \citenamefont {Samanta},\ and\ \citenamefont {Ellis}}]{Scuseria:2011ig}%
  \BibitemOpen
  \bibfield  {author} {\bibinfo {author} {\bibfnamefont {G.~E.}\ \bibnamefont
  {Scuseria}}, \bibinfo {author} {\bibfnamefont {C.~A.}\ \bibnamefont
  {Jim{\'e}nez-Hoyos}}, \bibinfo {author} {\bibfnamefont {T.~M.}\ \bibnamefont
  {Henderson}}, \bibinfo {author} {\bibfnamefont {K.}~\bibnamefont {Samanta}},
  \ and\ \bibinfo {author} {\bibfnamefont {J.~K.}\ \bibnamefont {Ellis}},\
  }\href@noop {} {\bibfield  {journal} {\bibinfo  {journal} {J. Chem. Phys.}\
  }\textbf {\bibinfo {volume} {135}},\ \bibinfo {pages} {124108} (\bibinfo
  {year} {2011})}\BibitemShut {NoStop}%
\bibitem [{\citenamefont {Qiu}\ \emph {et~al.}(2016)\citenamefont {Qiu},
  \citenamefont {Henderson},\ and\ \citenamefont {Scuseria}}]{Qiu:2016cf}%
  \BibitemOpen
  \bibfield  {author} {\bibinfo {author} {\bibfnamefont {Y.}~\bibnamefont
  {Qiu}}, \bibinfo {author} {\bibfnamefont {T.~M.}\ \bibnamefont {Henderson}},
  \ and\ \bibinfo {author} {\bibfnamefont {G.~E.}\ \bibnamefont {Scuseria}},\
  }\href@noop {} {\bibfield  {journal} {\bibinfo  {journal} {J. Chem. Phys.}\
  }\textbf {\bibinfo {volume} {145}},\ \bibinfo {pages} {111102} (\bibinfo
  {year} {2016})}\BibitemShut {NoStop}%
\bibitem [{\citenamefont {Jake}\ \emph {et~al.}(2018)\citenamefont {Jake},
  \citenamefont {Henderson},\ and\ \citenamefont {Scuseria}}]{Jake:2018ka}%
  \BibitemOpen
  \bibfield  {author} {\bibinfo {author} {\bibfnamefont {L.~C.}\ \bibnamefont
  {Jake}}, \bibinfo {author} {\bibfnamefont {T.~M.}\ \bibnamefont {Henderson}},
  \ and\ \bibinfo {author} {\bibfnamefont {G.~E.}\ \bibnamefont {Scuseria}},\
  }\href@noop {} {\bibfield  {journal} {\bibinfo  {journal} {J. Chem. Phys.}\
  }\textbf {\bibinfo {volume} {148}},\ \bibinfo {pages} {024109} (\bibinfo
  {year} {2018})}\BibitemShut {NoStop}%
\bibitem [{\citenamefont {Qiu}\ \emph {et~al.}(2018)\citenamefont {Qiu},
  \citenamefont {Henderson}, \citenamefont {Zhao},\ and\ \citenamefont
  {Scuseria}}]{Qiu:2018ei}%
  \BibitemOpen
  \bibfield  {author} {\bibinfo {author} {\bibfnamefont {Y.}~\bibnamefont
  {Qiu}}, \bibinfo {author} {\bibfnamefont {T.~M.}\ \bibnamefont {Henderson}},
  \bibinfo {author} {\bibfnamefont {J.}~\bibnamefont {Zhao}}, \ and\ \bibinfo
  {author} {\bibfnamefont {G.~E.}\ \bibnamefont {Scuseria}},\ }\href@noop {}
  {\bibfield  {journal} {\bibinfo  {journal} {J. Chem. Phys.}\ }\textbf
  {\bibinfo {volume} {149}},\ \bibinfo {pages} {164108} (\bibinfo {year}
  {2018})}\BibitemShut {NoStop}%
\bibitem [{\citenamefont {Fuchs}\ and\ \citenamefont
  {Schweigert}(1997)}]{Fuchs}%
  \BibitemOpen
  \bibfield  {author} {\bibinfo {author} {\bibfnamefont {J.}~\bibnamefont
  {Fuchs}}\ and\ \bibinfo {author} {\bibfnamefont {C.}~\bibnamefont
  {Schweigert}},\ }\href@noop {} {\emph {\bibinfo {title} {{Symmetries, Lie
  Algebras, and Representations}}}}\ (\bibinfo  {publisher} {Cambridge
  University Press},\ \bibinfo {year} {1997})\BibitemShut {NoStop}%
\bibitem [{\citenamefont {Gilmore}(2008)}]{Gilmore:2008}%
  \BibitemOpen
  \bibfield  {author} {\bibinfo {author} {\bibfnamefont {R.}~\bibnamefont
  {Gilmore}},\ }\href@noop {} {\emph {\bibinfo {title} {{Lie Groups, Physics,
  and Geometry: An Introduction for Physicists, Engineers and Chemists}}}}\
  (\bibinfo  {publisher} {Cambridge University Press},\ \bibinfo {year}
  {2008})\BibitemShut {NoStop}%
\bibitem [{Note1()}]{Note1}%
  \BibitemOpen
  \bibinfo {note} {Note that formally powers of $\protect \mathaccentV
  {hat}05EO_i^n$ are not defined in the Lie algebra, therefore, it is not
  possible to build Taylor series expandable functions for the operators
  $f(\protect \mathaccentV {hat}05EO_i)$, such functions would also commute
  with $\protect \mathaccentV {hat}05EH$ and could be added to the Lie algebra
  based on the commutation property. Products $\protect \mathaccentV
  {hat}05EO_i^n$ are only defined in the universal enveloping Lie
  algebra.}\BibitemShut {Stop}%
\bibitem [{\citenamefont {Barut}\ and\ \citenamefont {Raczka}(1980)}]{Barut}%
  \BibitemOpen
  \bibfield  {author} {\bibinfo {author} {\bibfnamefont {A.~O.}\ \bibnamefont
  {Barut}}\ and\ \bibinfo {author} {\bibfnamefont {R.}~\bibnamefont {Raczka}},\
  }\href@noop {} {\emph {\bibinfo {title} {{Theory of Group Representations and
  Applications}}}}\ (\bibinfo  {publisher} {Polish Scientific Publisher},\
  \bibinfo {year} {1980})\BibitemShut {NoStop}%
\bibitem [{\citenamefont {Wigner}(1959)}]{Wigner}%
  \BibitemOpen
  \bibfield  {author} {\bibinfo {author} {\bibfnamefont {E.~P.}\ \bibnamefont
  {Wigner}},\ }\href@noop {} {\emph {\bibinfo {title} {{Group theory and its
  application to the quantum mechanics of atomic spectra}}}}\ (\bibinfo
  {publisher} {Academic Press Inc.},\ \bibinfo {year} {1959})\BibitemShut
  {NoStop}%
\bibitem [{Note2()}]{Note2}%
  \BibitemOpen
  \bibinfo {note} {Construction of irreducible representations for other types
  of Lie algebras, such as abelian and reductive (direct sums of abelian and
  simple), is somewhat simpler because they can be built combining
  representations found for simple algebras and/or one-dimensional
  representations for abelian algebras.}\BibitemShut {Stop}%
\bibitem [{\citenamefont {Press}\ \emph {et~al.}(1992)\citenamefont {Press},
  \citenamefont {Flannery}, \citenamefont {Teukolsky},\ and\ \citenamefont
  {Vetterling}}]{num_rec}%
  \BibitemOpen
  \bibfield  {author} {\bibinfo {author} {\bibfnamefont {W.~H.}\ \bibnamefont
  {Press}}, \bibinfo {author} {\bibfnamefont {B.~P.}\ \bibnamefont {Flannery}},
  \bibinfo {author} {\bibfnamefont {S.~A.}\ \bibnamefont {Teukolsky}}, \ and\
  \bibinfo {author} {\bibfnamefont {W.~T.}\ \bibnamefont {Vetterling}},\
  }\href@noop {} {\emph {\bibinfo {title} {Numerical Recipes}}}\ (\bibinfo
  {publisher} {Cambridge University Press},\ \bibinfo {address} {Cambridge},\
  \bibinfo {year} {1992})\BibitemShut {NoStop}%
\bibitem [{\citenamefont {L\"owdin}(1964)}]{Lowdin}%
  \BibitemOpen
  \bibfield  {author} {\bibinfo {author} {\bibfnamefont {P.-O.}\ \bibnamefont
  {L\"owdin}},\ }\href@noop {} {\bibfield  {journal} {\bibinfo  {journal} {Rev.
  Mod. Phys.}\ }\textbf {\bibinfo {volume} {36}},\ \bibinfo {pages} {966}
  (\bibinfo {year} {1964})}\BibitemShut {NoStop}%
\bibitem [{\citenamefont {L\"owdin}(1955)}]{Lowdin3}%
  \BibitemOpen
  \bibfield  {author} {\bibinfo {author} {\bibfnamefont {P.-O.}\ \bibnamefont
  {L\"owdin}},\ }\href@noop {} {\bibfield  {journal} {\bibinfo  {journal}
  {Phys. Rev.}\ }\textbf {\bibinfo {volume} {97}},\ \bibinfo {pages} {1509}
  (\bibinfo {year} {1955})}\BibitemShut {NoStop}%
\bibitem [{\citenamefont {Kato}(1966)}]{Kato}%
  \BibitemOpen
  \bibfield  {author} {\bibinfo {author} {\bibfnamefont {T.}~\bibnamefont
  {Kato}},\ }\href@noop {} {\emph {\bibinfo {title} {{Perturbation theory for
  linear operators}}}}\ (\bibinfo  {publisher} {Springer-Verlag},\ \bibinfo
  {year} {1966})\BibitemShut {NoStop}%
\bibitem [{\citenamefont {Messiah}(1961)}]{Messiah}%
  \BibitemOpen
  \bibfield  {author} {\bibinfo {author} {\bibfnamefont {A.}~\bibnamefont
  {Messiah}},\ }\href@noop {} {\emph {\bibinfo {title} {{Quantum mechanics, Vol
  II }}}}\ (\bibinfo  {publisher} {North-Holland Publishing Company},\ \bibinfo
  {year} {1961})\BibitemShut {NoStop}%
\bibitem [{\citenamefont {{Gel'fand, Izrail' Moiseevich}}(1961)}]{Gelfand}%
  \BibitemOpen
  \bibfield  {author} {\bibinfo {author} {\bibnamefont {{Gel'fand, Izrail'
  Moiseevich}}},\ }\href@noop {} {\emph {\bibinfo {title} {{Lectures on Linear
  Algebra}}}}\ (\bibinfo  {publisher} {Interscience Publishers, Inc., New
  York},\ \bibinfo {year} {1961})\BibitemShut {NoStop}%
\end{thebibliography}
%

\appendix

\section*{Appendix A: Constructing projectors for infinite groups}

Symmetry operators $\{\hat O_k\}$'s commuting with $\hat H$ form Lie algebra $\mathcal{L}$:
\bea\label{eq:ACL}
[\hat O_1,\hat O_2] &=& \sum_k c_{12}^{(k)} \hat O_k,~ \hat O_k \in \mathcal{L} \\ 
a\hat O_1+b\hat O_2 &\in& \mathcal{L}, ~a,b\in \mathbb{C} 
\eea
Let us assume that we have $K$ linear independent elements of 
$\mathcal{L}$. There is a simple way to organize a compact Lie group $\GG$
(which is an infinite continuous group):
\bea 
g(\phi_1,...\phi_K) &=& \prod_{k=1}^{K} e^{i\phi_k \hat O_k}\in \GG, ~\phi_k\in[0,2\pi).
\eea
It is possible to verify the group axioms using the closure relation for the algebra $\LL$ (\eq{eq:ACL}).  
Of course, due to non-commutativity there are many ways to parametrize $\GG$, 
two simple alternatives can be 
\bea 
g'(\phi_1,...\phi_K) &=&  \exp\left[\sum_{k=1}^{K} i\phi_k \hat O_k\right], \\ \label{eq:EU}
g''(\phi_1,...\phi_K) &=&  e^{i\phi_K \hat O_{K-1}}\prod_{k=1}^{K-1} e^{i\phi_k \hat O_k},
\eea
where in the last equation the parametrization effectively introduces $\hat O_K$ via 
commutations between all other $\hat O_k$'s. 
There is a standard expression for projectors on irreducible representations of a compact group
\bea \label{eq:PCG}
\hat P_\Gamma = \int d\bar{\phi} \chi_\Gamma(\bar{\phi})^{*} g(\bar{\phi}),
\eea
where $\bar{\phi} = (\phi_1,...\phi_K)$, $\chi_\Gamma(\bar{\phi})$ is the character of the $\Gamma$ 
irreducible representation, and $d\bar{\phi}$ is the Haar measure defined over the group domain. 

To illustrate \eq{eq:PCG} in action let us consider the SO(3) group where the group is parametrized using 
the Euler angles (this parametrization is similar to that in \eq{eq:EU})
\bea\label{eq:gP}
g(\alpha,\beta,\gamma) &=& e^{i\alpha \hat J_z} e^{i\beta \hat J_y} e^{i\gamma \hat J_z}, \\\notag
&&\alpha,~\gamma\in[0,2\pi), ~\beta\in[0,\pi]
\eea
and the projector to a particular $\hat J^2$ and $\hat J_z$ eigenstate is 
\bea
\hat P_{jm} = \frac{2j+1}{8\pi^2} \int d\Omega \bra{jm}g(\Omega)\ket{jm} g(\Omega)
\eea
where $\Omega = (\alpha,\beta,\gamma)$. This projector can be further simplified using the 
$g$ parametrization in \eq{eq:gP} and the relation $\hat J_z\ket{jm} = m\ket{jm}$
\bea
\hat P_{jm} = \left(j+\frac{1}{2}\right) \int_{0}^{\pi} d\beta \sin(\beta) \bra{jm}e^{i\beta \hat J_y}\ket{jm} e^{i\beta \hat J_y}.
\eea
The eigenstates $\ket{jm}$ can be found independently as the basis of irreducible representations 
for the Lie algebra using the highest weight theorem and the corresponding method.\cite{Barut} Thus,
knowledge of irreducible representations for the corresponding Lie algebra is essential element of 
constructing projectors via the Lie group exponential map.  


\section*{Appendix B: Commutation of projectors for commuting operators}

Commutation of projectors on eigen-subspaces for commuting operators is straightforward 
to show if we consider all such operators in the common eigenstate basis 
\bea \notag
\hat O_i &=& \sum_j o_j^{(i)} \sum_k \ket{\phi_{j,k}}\bra{\phi_{j,k}} \\
&=& \sum_j o_j^{(i)} \hat P_j^{(i)}, ~ i=1,2 
\eea
Here index $k$ enumerates eigen-basis states corresponding to the same eigenvalue.
Then, the commutator of the projector operators is 
\bea   \notag
 \left[ \hat P_1^{(j)},\hat P_2^{(l)} \right] &=&  
 \sum_{k,k'} \ket{\phi_{j,k}}\bra{\phi_{j,k}}\phi_{l,k'}\rangle\bra{\phi_{l,k'}} \\
 &-& \ket{\phi_{l,k'}}\bra{\phi_{l,k'}}\phi_{j,k}\rangle\bra{\phi_{j,k}}= 0.
\eea
The last equality is a consequence of the equalities for inner products 
$\bra{\phi_{l,k'}}\phi_{j,k}\rangle=\delta_{kk'}\delta_{jl}$
and $\bra{\phi_{j,k}}\phi_{l,k'}\rangle = \delta_{kk'}\delta_{jl}$.
In the case when these inner products are 0, $\hat P_1^{(j)}\hat P_2^{(l)}\equiv 0$ thus pairing such 
projectors will not give rise to non-trivial operators. For example, if the product of 
two projectors is $\hat P_{S^2=0} \hat P_{S_z=1}$, it clearly is $\equiv 0$ because this combination 
violates the usual conditions on the 
ranges of eigenvalues of $\hat S^2$ and $\hat S_z$ operators.
Thus, knowledge of irreducible representations of the corresponding 
Lie algebra is crucial to avoid pairings of projectors that produce trivial operators.

\end{document}